\documentclass[dvips,reprint,twocolumn]{revtex4-1}
\usepackage{amssymb,amsfonts,amsmath}
\usepackage{graphicx}

\begin{document}
\renewcommand{\thefootnote}{\fnsymbol{footnote}}
\title{Preparation of Nuclear Spin Singlet States using Spin-Lock Induced Crossing}

\author{Stephen J. DeVience}
\email{devience@fas.harvard.edu}
\affiliation{Department of Chemistry and Chemical Biology, Harvard University, 12 Oxford St., Cambridge, MA 02138, USA.}
 
\author{Ronald L. Walsworth}
\email{rwalsworth@cfa.harvard.edu}
\affiliation{Harvard-Smithsonian Center for Astrophysics, 60 Garden St., Cambridge, MA 02138, USA.}
\affiliation{Center for Brain Science, Harvard University, 52 Oxford St., Cambridge, MA 02138, USA.}
\affiliation{Department of Physics, Harvard University, 17 Oxford St., Cambridge, MA 02138, USA.}

\author{Matthew S. Rosen}
\email{mrosen@cfa.harvard.edu}
\affiliation{Department of Physics, Harvard University, 17 Oxford St., Cambridge, MA 02138, USA.}
\affiliation{Harvard Medical School, 25 Shattuck Street, Boston, MA 02115, USA.}
\affiliation{A. A. Martinos Center for Biomedical Imaging, 149 Thirteenth St., Charlestown, MA 02129, USA.}

\begin{abstract}
We introduce a broadly applicable technique to create nuclear spin singlet states in organic molecules and other many-atom systems. We employ a novel pulse sequence to produce a spin-lock induced crossing (SLIC) of the spin singlet and triplet energy levels, which enables triplet/singlet polarization transfer and singlet state preparation. We demonstrate the utility of the SLIC method by producing a long-lived nuclear spin singlet state on two strongly-coupled proton pairs in the tripeptide molecule phenylalanine-glycine-glycine dissolved in D$_2$O, and by using SLIC to measure the J-couplings, chemical shift differences, and singlet lifetimes of the proton pairs. We show that SLIC is more efficient at creating nearly-equivalent nuclear spin singlet states than previous pulse sequence techniques, especially when triplet/singlet polarization transfer occurs on the same timescale as spin-lattice relaxation. 
\end{abstract}

\maketitle

There is great current interest in the controlled preparation and coherent manipulation of singlet states for nuclear spin pairs in molecules and other many-atom systems (e.g., spin networks in solids), as spin singlet states are largely decoupled from environmental perturbations that limit the spin state lifetime. For example, in liquid state experiments singlet states in nuclear spin pairs can exhibit lifetimes much longer than the single-spin polarization lifetime ($T_1$) \cite{Levitt1, Levitt5, Levitt6, Levitt7, Levitt11, Ghosh1, Bodenhausen4, Bodenhausen5, DeVience1, Pileio3, Warren1, Warren2}. In addition, nuclear spin singlet states can be used as a resource for spectroscopic interrogation of couplings within many-spin systems, including J-couplings, dipolar, and hyperfine couplings in both organic molecules and spin networks in solids \cite{Levitt2, Levitt8, Levitt10}. Such singlet states exist naturally when nuclear spins are strongly J-coupled relative to their resonance frequency differences, $\Delta \nu$, i.e., $J >> \Delta \nu$. However, due to the differences in spin singlet and triplet state symmetries, it is not possible to transfer polarization from the triplet to the singlet state by directly driving a radiofrequency transition, which limits the control of singlet state preparation and manipulation. Tayler and Levitt demonstrated that triplet/singlet polarization transfer can instead be acheived using a series of $\pi$-pulse trains in which the pulse timing is synchronized to the J-coupling strength between nuclei \cite{Levitt11}. This ``M2S'' sequence takes advantage of the small amount of mixing between singlet and triplet states that is present whenever $\Delta \nu > 0$. Feng and Warren also showed that the M2S sequence can create singlet states in certain heteronuclear systems even when $\Delta \nu = 0 $ \cite{Warren2}. These results hold promise for creating hyperpolarized singlet states without the need for a symmetry-breaking chemical reaction or continuous spin-locking \cite{Warren1}. However, in all results to date, the polarization transfer to the spin singlet state only occurs during the final third of the M2S sequence time, and before this stage the spin polarization occupies states subject to conventional spin-lattice relaxation. 

In this letter, we show that better triplet/singlet polarization transfer efficiency can be acheived by replacing the M2S pulse trains with a continuous spin-lock whose nutation frequency is matched to the J-coupling between the target nuclear spins. At this spin-locking strength, the energy levels of the singlet state and one triplet state become equal in the rotating frame, which we call the ``spin-lock induced crossing'' (SLIC). Polarization transfer can occur for the duration of spin-locking, which minimizes polarization loss from triplet state relaxation and thus provides better efficiency for singlet state creation than the M2S technique. SLIC is analogous to the Hartmann-Hahn condition for polarization transfer between two magnetically inequivalent nuclear spins, except that for SLIC the nuclei are nearly identical and their spin symmetry subspaces are inequivalent \cite{Hartmann}. 

We experimentally compare M2S and SLIC using liquid-state NMR of the tripeptide molecule phenylalanine-glycine-glycine (phe-gly-gly), which contains two nearly-equivalent proton spin pairs in which to prepare singlet states. We find that for these two proton spin pairs, singlet state creation with SLIC is 19\% and 75\% more efficient than with M2S. We also demonstrate the utility of SLIC for characterizing singlet state lifetimes as well as small J-couplings and chemical shift differences between nearly-identical nuclear spins.

\begin{figure*}
\vspace*{.05in}
\centering
\includegraphics[width=6.86in]{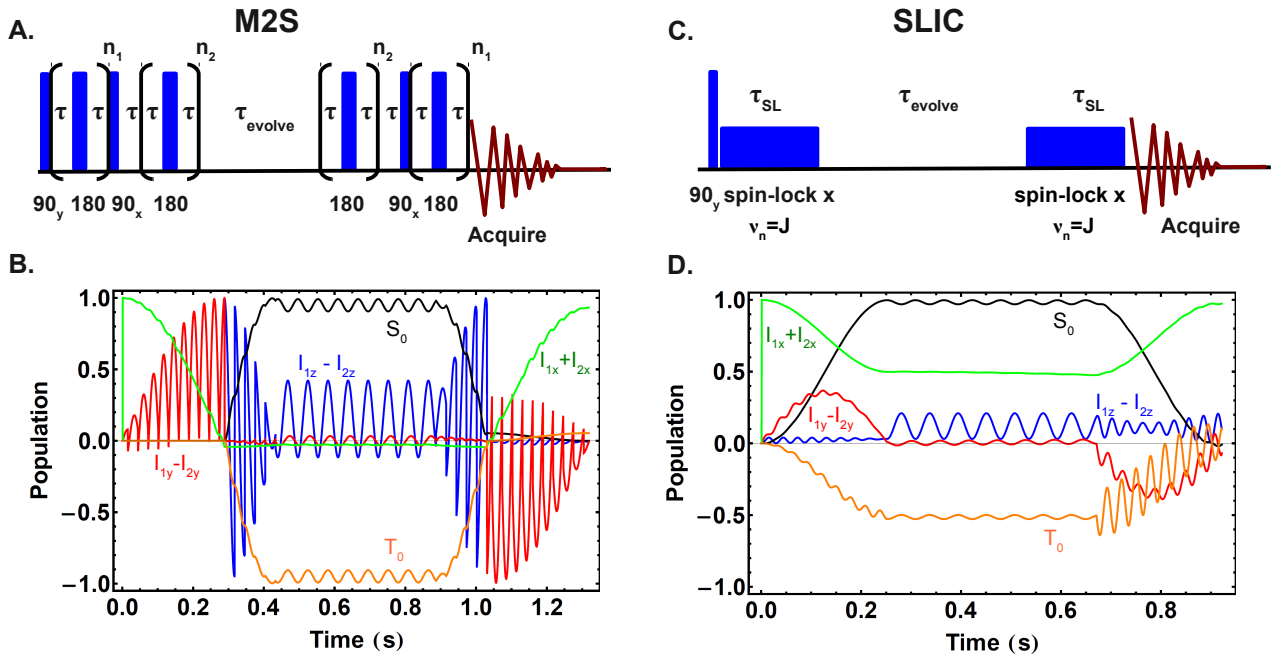} 
\caption{Simulated comparison of M2S and SLIC techniques applied to singlet state creation,  evolution, and readout for a proton spin pair with chemical shift $\delta$ = 3.71 ppm in the phe-gly-gly molecule. (A) Schematic of M2S experiment: Following an initial 90 degree pulse to create transverse triplet polarization, the M2S pulse sequence creates singlet state population by applying pulse trains with appropriate length and pulse spacing synchronized with the proton pair's J-coupling and resonance frequency difference. The system then evolves over $\tau_{evolve}$, and the M2S sequence is applied in reverse order to convert singlet population back to transverse triplet polarization for inductive NMR detection. (B) Simulation of M2S experiment: Transfer, in a series of stages, of transverse (x-axis) triplet polarization to singlet-triplet coherences and finally to singlet and triplet state populations of equal magnitudes. Transfer to the singlet state population only occurs in the final 1/3 of the M2S preparation sequence. (C) Schematic of SLIC experiment: Transverse triplet polarization is created via an initial 90 degree pulse and is then transferred to singlet state population by the application of a spin-locking field with $\nu_n = J$ for period $\tau_{SL}$. The system then evolves for $\tau_{evolve}$ and an identical spin-locking field converts singlet population back to transverse triplet polarization for detection. (D) Simulation of SLIC experiment: Transfer of transverse triplet polarization to singlet state population begins immediately and occurs during a single spin-locking stage.}
\end{figure*}

Figure 1A shows the M2S experimental protocol used to create a nuclear spin singlet state from triplet-state polarization and then return the singlet state to transverse triplet-state (i.e., measurable) polarization after an evolution time, $\tau_{evolve}$. Figure 1B gives a simulation of spin state and coherence dynamics during singlet state preparation with M2S if relaxation is ignored \cite{spindynamica}. The first pulse train in Fig. 1A converts the triplet-state polarization into a singlet-triplet coherence with a relaxation time of $T_2/3 \approx T_1/3$ (for liquid-state NMR of small molecules), and the second pulse train creates a singlet population with relaxation time $T_S$ \cite{DeVience1, Bodenhausen1}. The number of pulses required for the M2S sequence increases as the resonance frequency difference ($\Delta \nu$) between the two nearly-identical nuclear spins decreases and hence the singlet state becomes closer to ideal. In many cases, the required M2S pulse sequence time approaches or exceeds $T_1$ of the nuclear spins, and significant spin polarization can be lost before it is transferred to the singlet state, particularly during the first 2/3 of the sequence. For an ideal system, the time required for maximum singlet state creation is 

\begin{equation}
t_{M2S,max} \approx \frac{3 \pi}{8 \Delta \nu} = \frac{1.18}{\Delta \nu}
\end{equation}

Figure 1C shows the SLIC pulse sequence used to create a nuclear spin singlet state from triplet-state polarization and return the singlet state to transverse triplet-state polarization after an evolution time ($\tau_{evolve}$) in analogy to the M2S experiment. However, instead of pulse trains, continuous spin-locking is applied at a nutation frequency equal to the J-coupling between spins, i.e., $\nu_n = J$. The simulation shown in Fig. 1D illustrates that such spin-locking transfers triplet-state polarization directly from transverse polarization into singlet state population more quickly than in the M2S sequence. A density matrix analysis (detailed in the supplement) shows that selecting a nutation frequency $\nu_n = J$ matches the energies of the singlet state and one of the triplet states, creating a spin-lock induced crossing. At this energy, off-diagonal interaction terms $\Delta \nu / 2\sqrt{2}$ become significant and induce oscillatory triplet/singlet polarization transfer with a period of $\sqrt{2}/\Delta\nu$ and maximimum transfer to the singlet state at half this time:

\begin{equation}
t_{SL,max} = \frac{1}{\Delta \nu \sqrt{2}} = \frac{0.707}{\Delta \nu}  .
\end{equation}

Comparison with equation 1 shows that SLIC produces singlet state polarization about 40\% faster than M2S, which results in fewer relaxation losses. To compare the effectiveness of the two sequences, we performed simulations using Bloch equations to model relaxation and singlet/triplet polarization transfer. M2S was modeled in two steps: first, polarization transfer from $I_{1x}+I_{2x}$ with lifetime $T_2 = T_1$ to $I_{1y}-I_{2y}$ with lifetime $T_1/3$; second, polarization transfer from $I_{1z}-I_{2z}$ with lifetime $T_1/3$ to singlet state $S_0$ with lifetime $T_S$. Only one polarization transfer needed to be modeled for SLIC, between $I_{1x}+I_{2x}$ with lifetime $T_2 = T_1$ and $S_0$ with lifetime $T_S$. A maximum of 50\% polarization transfer to the singlet state can be achieved by both sequences, which we define to be an efficiency of 100\%. Figure 2 plots the calculated polarization transfer efficiency for M2S and SLIC as a function of the product $T_1 \Delta \nu$. Two cases are considered, one in which $T_S >> T_1$, and one in which $T_S = 3 T_1$. SLIC is found to be significantly more efficient than M2S for all ranges of parameters, and particularly for $T_1 \Delta \nu < 1$. Note that both sequences are less efficient for smaller $T_S / T_1$ due to singlet relaxation.

\begin{figure}
\vspace*{.05in}
\centering
\includegraphics[width=3.43in]{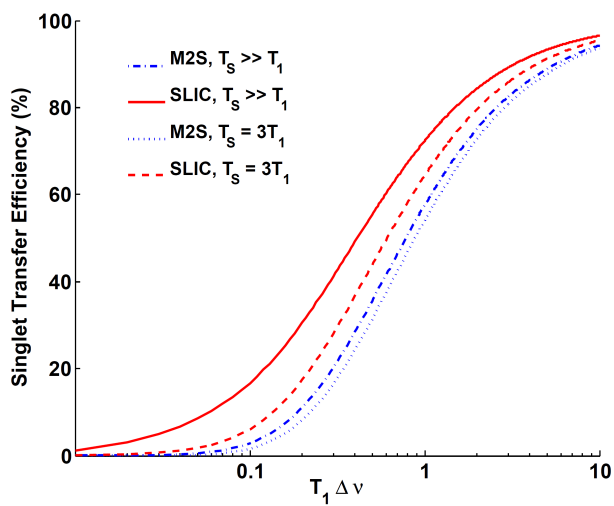}  
\caption{Simulations of ideal triplet/singlet polarization transfer efficiency for M2S (blue) and SLIC (red). Results are shown for $T_S >> T_1$ and $T_S = 3 T_1$.}
\end{figure}

To assess the relative utility of the SLIC and M2S sequences for producing nuclear spin singlet states, we performed NMR measurements at 4.7 T on a 20 mM solution of the phe-gly-gly molecule dissolved in D$_2$O, addressing a nearly-equivalent proton spin pair with chemical shift $\delta$ = 3.71 ppm and $T_1$ = 912 $\pm$ 7 ms. For the M2S sequence (Fig. 1A) we found optimized parameters for singlet creation to be n$_1$ = 10, n$_2$ = 5, and $\tau$ = 14.4 ms, which indicates $J = 17.4\pm 0.1$ Hz and $\Delta \nu = 2.8 \pm 0.3$ Hz. We also found a singlet lifetime of T$_S$ = 25.1 $\pm$ 0.8 s with no spin-locking applied during $\tau_{evolve}$. We measured the NMR signal intensity (x-axis magnetization,proportional to the transverse triplet-state polarization) at the end of the M2S sequence for $\tau_{evolve}$ = 5 s, which arises from the transfer of transverse triplet-state polarization to singlet-state population and then back to measurable transverse triplet-state polarization after $\tau_{evolve}$, and we then used the singlet lifetime to extrapolate the singlet-state population at $\tau_{evolve}$ = 0. During $\tau_{evolve}$, remaining triplet-state polarization is lost to relaxation, and it does not contribute to the final signal. We compared this M2S NMR signal magnitude to a reference measurement arising from a single 90 degree pulse applied to the sample, i.e., without singlet creation. From this analysis we estimate that 24\% of the initial triplet-state polarization was transferred to the singlet state and back to triplet for $\tau_{evolve}$ = 0, out of a theoretical maximum of 50\%, yielding an efficiency of 69\% for each application of the M2S sequence. 

\begin{figure*}
\vspace*{.05in}
\centering
\includegraphics[width=6.86in]{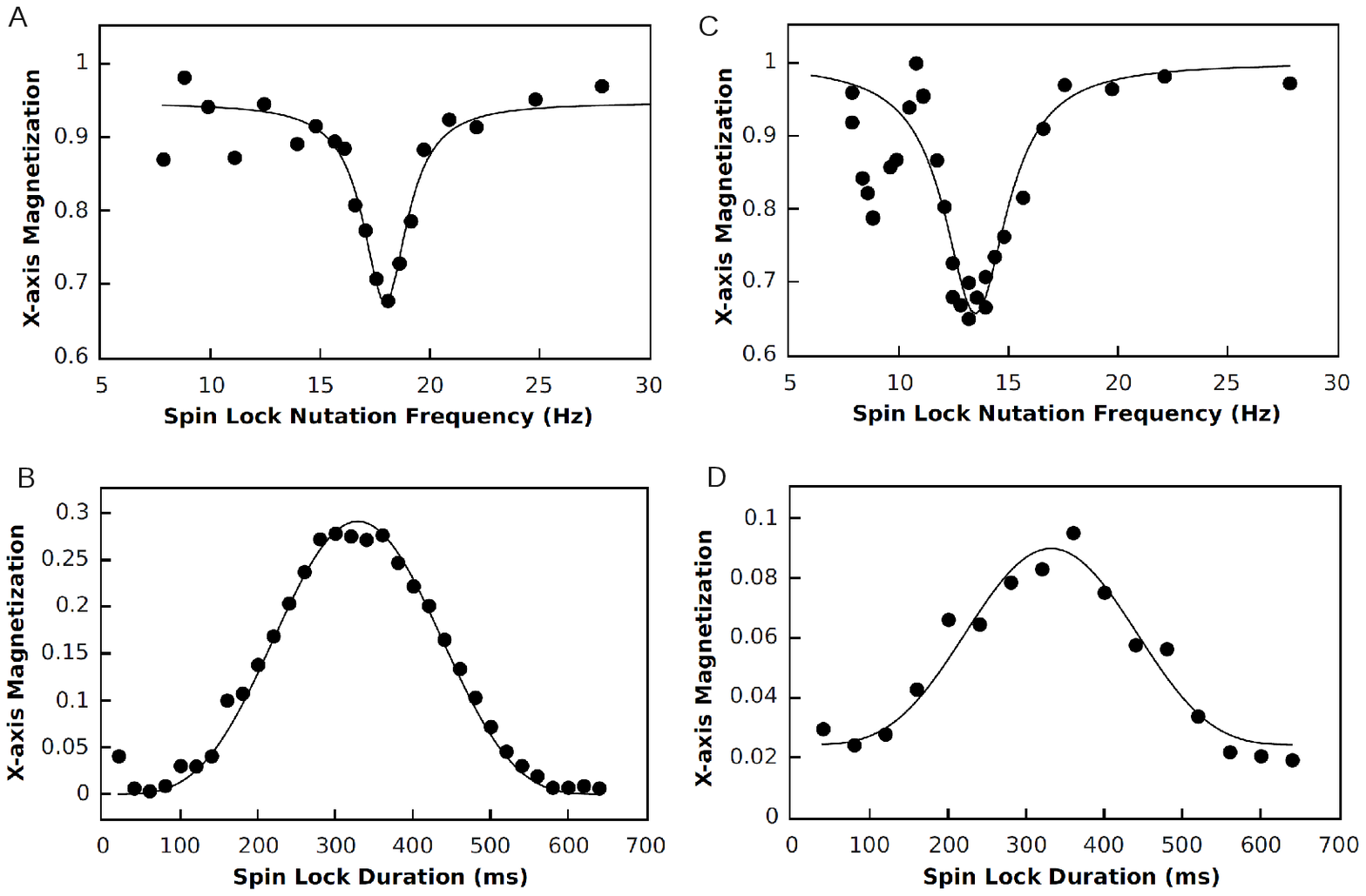}  
\caption{Experimental application of the SLIC technique to nuclear spin singlet state creation in the phe-gly-gly molecule. Results for $\delta$ = 3.71 ppm proton pair: (A) The NMR signal (normalized x-axis magnetization, proportional to transverse triplet polarization) following the first SLIC spin-lock as a function of nutation frequency for $\tau_{SL}$ = 300 ms exhibits a pronounced dip when the spin-lock frequency equals the J-coupling. A Lorentzian is fit to the measurement to determine $\nu_n =J= 17.5 \pm 0.3$ Hz. (B) NMR signal after the complete SLIC experiment with $\tau_{evolve}$ = 5 s (proportional to final transverse triplet polarization surviving transfer to singlet and back) as a function of spin-lock duration. Maximal singlet state creation is found for $\tau_{SL} \approx$ 280 to 360 ms. From a fit with the function sin$^4 (2 \pi \tau_{SL}/T)$, we find $\Delta \nu$ = 2.15 $\pm$ 0.01 Hz. Results for $\delta$ = 3.20 ppm proton pair: (C) NMR signal following the first SLIC spin-lock as a function of nutation frequency for $\tau_{SL}$ = 332 ms, with a Lorentzian fit to the data yielding optimal $\nu_n = J = 13.5 \pm 0.2$ Hz. (D) NMR signal after the complete SLIC experiment with $\tau_{evolve}$ = 500 ms as a function of spin-lock duration. Maximal singlet state creation is found for $\tau_{SL} \approx$ 300 to 400 ms. From a fit with the function sin$^4 (2 \pi \tau_{SL}/T) + c$, we find $\Delta \nu$ = 2.13 $\pm$ 0.04 Hz.}
\end{figure*}

For the SLIC technique, we determined the optimal spin-lock nutation frequency by performing a truncated SLIC pulse sequence in which the NMR signal (x-axis magnetization) was acquired directly after the first spin-locking period. As a function of nutation frequency, the measured NMR signal exhibited a dip centered at $\nu_n$ = 17.5 $\pm$ 0.3 Hz with a relative depth of $\approx 25\%$ (Fig. 3A), consistent with the SLIC condition of $\nu_n = J$ for optimal triplet/singlet polarization transfer. We then used this optimal spin-lock nutation frequency in the complete SLIC sequence with $\tau_{evolve}$ = 5 s (Fig. 1C) and optimized the spin-lock duration ($\tau_{SL}$) to produce the strongest NMR signal and hence maximal singlet state creation. The measured dependence of the SLIC NMR signal on the spin-lock duration (Fig. 3B) exhibits a flat maximum for $\tau_{SL} \approx$ 280 to 360 ms. Using $\tau_{SL}$ = 300 ms provided about 34\% polarization transfer from the triplet to the singlet state and back when extrapolated to $\tau_{evolve}$ = 0, indicating an 82\% polarization transfer efficiency for each application of SLIC spin-locking.

We next applied the SLIC technique to singlet state creation in a second proton spin pair in the phe-gly-gly molecule, with chemical shift $\delta$ = 3.20 ppm and $T_1$ = 430 $\pm$ 5 ms. This proton spin pair is coupled to a third proton that decreases both the singlet lifetime ($T_S$ = 2.15$\pm$ 0.05 s with no spin-locking applied) and the triplet/singlet polarization transfer efficiency. We followed the procedure outlined above to determine the optimal spin-lock nutation frequency $\nu_n = J = 13.5 \pm 0.2$ Hz, and the optimal spin-lock duration $\tau_{SL}$ = 332 $\pm$ 6 ms (Fig. 3C,D). We then applied the complete SLIC sequence with $\tau_{evolve}$ = 500 ms and found about 12\% polarization transfer to the singlet state and back when extrapolated to $\tau_{evolve}$ = 0, which represents a transfer efficiency of 49\% for each application of SLIC spin-locking. For comparison, we experimentally investigated singlet state creation with the M2S sequence. We determined optimized M2S parameters to be n$_1$ = 4, n$_2$ = 5, and $\tau$ = 17.9 ms, and measured that only 4\% of the polarization was transferred from the triplet to the singlet state and back when extrapolated to $\tau_{evolve}$ = 0, which represents a 28\% efficiency for each application of the M2S sequence. 

In summary, we introduced an improved and broadly applicable method, known as SLIC for ``spin-lock induced crossing,'' for the creation of long-lived singlet states of nuclear spins in molecules and other many-atom systems.  As an example, we applied our SLIC technique to two different nearly-equivalent proton spin pairs in the phe-gly-gly molecule and demonstrated that SLIC is 40\% and 300\% more efficient than the previous M2S technique for the transfer of triplet-state polarization to singlet-state population and then back to measurable transverse triplet-state polarization. SLIC is more effective than M2S primarily because the transfer to the long-lived singlet state begins immediately with SLIC, without the need for an initial transfer to a singlet-triplet coherence as with M2S. Though a singlet-triplet coherence can have an extended lifetime relative to a single-spin coherence time ($T_2$), it generally relaxes significantly faster than the singlet population lifetime ($T_S$), leading to greater polarization loss and less efficient singlet state creation for M2S than for SLIC. The relative advantage of SLIC grows for molecules with small resonance frequency difference $\Delta \nu$ between the nuclear spins, which in many cases can be much smaller than 1/$T_1$. These are the very molecules expected to possess the longest singlet lifetimes due to the high purity of their singlet states.

Beyond liquid-state NMR, we foresee applications of SLIC in long-lived quantum memories composed of nuclear spin pairs \cite{Maurer1}, as well as selective nuclear spin state manipulation at low magnetic field \cite{DeVience2}, and high-precision solid-state electronic spin measurements such as nitrogen vacancy (NV) diamond magnetometry \cite{Walsworth1,Rugar1,Wrachtrup1,Zhao1}. For a nuclear spin quantum memory with singlet and triplet bases, SLIC may provide a straightforward way to prepare desired states on the singlet/triplet Bloch sphere.  At low magnetic field, nuclear spins are strongly coupled and do not exhibit resolvable spectral features for manipulation using conventional single-spin NMR techniques. However, SLIC could be used to mix and thereby manipulate nuclear spin states by exploiting their weak chemical shift differences. SLIC may also provide a tool for improving the sensitivity of NV diamond magnetometry by allowing state-specific preparation and manipulation of the nearly-equivalent nitrogen electron spins that occur in the diamond lattice at about ten times the concentration of optically polarizable and detectable NV centers. These nitrogen electron spins could be prepared in singlet states to increase the NV coherence time, or in specific triplet states to serve as ancillary magnetic field sensors controllably coupled via dressed-state techniques to the optically detectable NV centers \cite{Cappellaro1, Lukin1}.

\bibliographystyle{apsrev4-1}
\bibliography{singlet_bib}

\end{document}